\documentclass{emulateapj}

\newcommand{\metal}{[Fe/{}H]}
\newcommand{\oh}{[O/{}H]}
\newcommand{\cfe}{[C/{}Fe]}
\newcommand{\nfe}{[N/{}Fe]}
\newcommand{\ofe}{[O/{}Fe]}
\newcommand{\ch}{[C/{}H]}
\newcommand{\nh}{[N/{}H]}

\newcommand{\co}{{CO}}

\newcommand{\cn}{{CN}}

\newcommand{\cemp}{{CEMP}}
\newcommand{\ctwo}{{C$_2$}}

\newcommand{\teff}{{$T_{\rm eff}$}}
\newcommand{\logg}{{log \textit{g}}}
\newcommand{\cfeh}{{\cfe{}$_{h}$}}
\newcommand{\nfeh}{{\nfe{}$_{h}$}}
\newcommand{\sigofe}{{$\sigma_{\ofe{}}$}}

\begin{document}

\title{\ofe{} Estimates for Carbon-Enhanced Metal-Poor Stars from Near-IR Spectroscopy}

\author{Catherine R. Kennedy}
\affil{Department of Physics \& Astronomy and JINA: Joint Institute for Nuclear Astrophysics, Michigan State
University, \\ East Lansing, MI 48824, USA}
\email{kenne257@msu.edu}

\author{Thirupathi Sivarani}
\affil{Indian Institute of Astrophysics, \\ 2nd Block, Koramangala, Bangalore 560034, India}
\email{sivarani@iiap.res.in}

\author{Timothy C. Beers, Young Sun Lee}
\affil{Department of Physics \& Astronomy and JINA: Joint Institute for Nuclear Astrophysics, Michigan State
University, \\ East Lansing, MI 48824, USA}
\email{beers@pa.msu.edu, leeyou25@msu.edu}

\author{Vinicius M. Placco, Silvia Rossi}
\affil{Departamento de Astronomia - Instituto de Astronomia, Geof\'isica e Ci\^encias Atmosf\'ericas, Universidade de S\~ao Paulo, S\~ao Paulo, SP 05508-900, Brazil}
\email{vmplacco@astro.iag.usp.br, rossi@astro.iag.usp.br}

\author{Norbert Christlieb}
\affil{Zentrum f\"ur Astronomie der Universit\"at Heidelberg, Landessternwarte, K\"onigstuhl 12, 69117, Heidelberg, Germany}
\email{N.Christlieb@lsw.uni-heidelberg.de}

\author{Falk Herwig}
\affil{Department of Physics \& Astronomy, University of Victoria, Victoria, BC V8P5C2 Canada}
\email{fherwig@uvic.ca}

\author{Bertrand Plez}
\affil{GRAAL-CNRS, Universit\'{e} Montpellier II - UMR 5024, Place Eug\`{e}ne Bataillon, F-34095, Montpellier Cedex 05, France}
\email{bertrand.plez@graal.univ-montp2.fr}

\accepted{for publication in AJ -- January 9, 2011}

\begin{abstract}

We report on oxygen abundances determined from medium-resolution near-IR
spectroscopy for a sample of 57 carbon-enhanced metal-poor (\cemp{}) stars
selected from the Hamburg/ESO survey. The majority of our program stars exhibit
oxygen-to-iron ratios in the range $+0.5<$ \ofe{}$ <+2.0$. The \ofe{} values for
this sample are statistically compared to available high-resolution estimates for known CEMP
stars, as well as to high-resolution estimates for a set of carbon-normal
metal-poor stars. Carbon, nitrogen, and oxygen abundance patterns for a
sub-sample of these stars are compared to yield predictions for
very metal-poor asymptotic giant-branch abundances in the recent literature. We
find that the majority of our sample exhibit patterns that are consistent with
previously studied CEMP stars having s-process-element enhancements, and thus
have very likely been polluted by carbon- and oxygen-enhanced material
transferred from a metal-poor asymptotic giant-branch companion.            

\end{abstract}

\section{Introduction}

Carbon-enhanced metal-poor (\cemp{}) stars are quite common in the halo
populations of the Milky Way, and are of particular interest, as they preserve
important astrophysical information concerning the early chemical evolution of
the Galaxy (Beers \& Christlieb 2005).  Previous work has indicated that at least 20\% of stars with metallicities
[Fe/H] $< -2.0$ exhibit large over-abundances of carbon ([C/Fe] > +1.0; Lucatello et al. 2006; Marsteller et al. 2009), although recent studies (e.g.
Cohen et al. 2005; Frebel et al. 2006), have claimed that this fraction is somewhat lower (9\% and
14\%, respectively, for [Fe/H] $< -2.0$). In any case, the fraction of CEMP
stars rises to 30\% for [Fe/H] $< -3.0$, 40\% for [Fe/H] $< -3.5$, and 100\% for
[Fe/H] $< -4.0$ (Beers \&
Christlieb 2005; Frebel et al. 2005; Norris et al. 2007).

There exist a number of classes of \cemp{} stars,
some of which have been associated with proposed progenitor objects. CEMP-s
stars (those with s-process-element enhancement), for example, are the most
commonly observed type to date. High-resolution spectroscopic studies have
revealed that around 80\% of \cemp{} stars exhibit s-process-element enhancement
(Aoki et al. 2007). The favored mechanism invoked to account for 
these stars is mass transfer of carbon-enhanced material from the envelope of an
asymptotic giant-branch (AGB) star to its binary companion; it is this surviving
binary companion that is now observed as a CEMP-s star. 

The class of CEMP-no stars (which exhibit no strong neutron-capture-element
enhancements) are particularly prevalent among the most metal-poor stars.
Possible progenitors for this class include massive, rapidly-rotating, mega
metal-poor ([Fe/H] $< -6.0$) stars, which models suggest have greatly enhanced
abundances of CNO due to distinctive internal burning and mixing episodes,
followed by strong mass loss (Meynet et al.
2006; Hirschi et al. 2006; Meynet et al. 2010). Another suggested
mechansim is pollution of the interstellar medium by so-called faint supernovae
associated with the first generations of stars, which experience extensive
mixing and fallback during their explosions (Umeda \&
Nomoto 2003, 2005; Tominaga et al. 2007); high [C/Fe] and [O/Fe] ratios are predicted in the ejected
material. This model well reproduces the observed abundance pattern of the
CEMP-no star BD+44:493, the ninth-magnitude [Fe/H] $= -3.7$ star (with [C/Fe]$ =
+1.3$, [N/Fe] $= +0.3$, [O/Fe] $= +1.6$) discussed by Ito et al. (2009).

The great majority of
known CEMP stars were originally identified as metal-poor candidates from
objective-prism surveys, such as the HK survey (Beers et al. 1985,
1992), and
the Hamburg/ESO Survey (HES; Christlieb
2003; Christlieb et al. 2008), based on a
weak (or absent) \ion{Ca}{2} K line. Some candidate CEMP stars also come from a list
of HES stars selected from the prism plates based on their strong molecular
lines of carbon (Christlieb et al. 2001). Medium-resolution spectra for most of
these objects have been obtained over the past few years (Goswami et al. 2006; Marsteller 2007;
Goswami et al. 2010, Sivarani et al., in preparation). Inspection
of these data indicate that at least 50\% of these targets are consistent with
identification as CEMP stars, while the others are roughly solar-metallicity
carbon-rich stars. Dedicated surveys for CEMP stars covering a wide range of
carbon abundance and metallicities are just now getting underway, based on the
observed strength of the CH G band measured from the HES prism plates (e.g.,
Placco et al. 2010).

In order to more fully test the association of CEMP-no stars with massive
primordial stars and/or faint supernovae, and to better explore the nature of
the s-process in low-metallicity AGB stars (which is still rather poorly
understood; Herwig 2005), we require measurements of the important elements
C, N, and O for as large a sample of CEMP stars as possible. While estimates of
carbon and nitrogen abundances can be determined from medium-resolution optical
or near-UV spectra of CEMP stars (e.g., Rossi et al. 2005; Beers et al. 2007b; Johnson
et al. 2007; Marsteller et al. 2009), high-resolution spectroscopy is usually
required in order to obtain estimates of oxygen abundances from the forbidden
[\ion{O}{1}] $\lambda$6300 \AA{} line, the $\lambda$7700 \AA{} triplet
(e.g., Schuler et al. 2006;
Sivarani et al. 2006; Fabbian et al. 2009, and references
therein), or the OH lines at 1.5-1.7 $\micron$ (Mel´endez \& Barbuy
2002). Masseron et al. (2010) provides a useful compilation of known elemental
abundances for CEMP stars. In addition to abundance measurements for metal-poor
halo stars, oxygen abundances have also been measured directly in the
gas phase in damped Lyman $\alpha$ systems (Pettini et al. 2002, 2008).

If a star has a measured carbon abundance (and, assuming C/O $> 1$, which
applies for most CEMP stars), essentially all of the O is locked up in CO
molecules, and medium-resolution spectroscopy of the \co{} ro-vibrational bands
in the near-IR can be used for estimation of \ofe{} (e.g., Beers et al. 2007b, and references therein). Although one sacrifices measurement accuracy,
relative to high-resolution studies, this approach has the great
advantage that medium-resolution spectroscopy can be gathered far faster than
high-resolution spectroscopy, ensuring that much larger samples of stars can be
investigated. In addition, the large separation of the $^{13}$CO lines from the
$^{12}$CO lines at 2.3$\mu$m provides a straightforward means to measure the
important mixing diagnostic $^{12}$C/$^{13}$C, as long as the S/N of the spectra
are sufficient. 

This paper is outlined as follows. In Section 2 we discuss details of the
observations and data reduction procedures used in the present study. Section 3
describes the previously determined atmospheric parameter estimates and their
origins, as well as details about the synthetic spectra.  Methods used for determination of \ofe{} for our sample of stars are
described in Section 4. Our results, and a statistical comparison to high-resolution
estimates of [C/Fe] and [O/Fe] for a subset of our program stars can be found in
Section 5. Section 6 is a short discussion of our results; conclusions follow in
Section 7.     

\section{Observations and Data Reduction}

Our sample of 57 stars was selected from the Hamburg/ESO Survey (HES), based on follow-up medium-resolution optical spectra obtained during the course of searches for low-metallicity stars.  These optical spectra were
obtained with the GOLDCAM spectrograph on the KPNO 2.1m telescope and with the RC
Spectrographs on the 4m KPNO and CTIO telescopes (see
Beers et al. 2007b, hereafter Paper I).  Additional targets were selected from the list of carbon-rich candidates published by Christlieb et al. (2001) with available optical spectra.  Based on the optical spectra, all of the candidates are metal-poor stars, spanning the metallicity range $-2.8 \le$ \metal{} $\le -1.0$.  All of the stars were selected to be carbon-rich, with the majority exhibiting \cfe{} $\ge +1.0$, and thus are carbon-enhanced as defined by Beers \& Christlieb (2005).  Since our intention was to obtain near-IR spectroscopy of the CO features, the stars were also selected to have effective temperatures less than 5000 K, since warmer stars do not exhibit strong CO.

Estimates of \ofe{} for our program stars are derived from analysis of
medium-resolution near-IR spectra taken with the SOAR 4.1m telescope, using the
OSIRIS (Ohio State InfraRed Imager/Spectrometer; Depoy et al.
1993)
spectrograph during October 2005 to June 2008.  We used the long slit (width set to 1\arcsec) and long camera (with focal ratio f/7),
which provided a resolving power $R = 3000$. The long-pass K-band filter was used
to isolate the spectral region from 2.25$\mu$m to 2.45$\mu$m. Visible in this
band are the four ro-vibrational \co{} features used for the determination of
\ofe{}. We also observed A0-type stars at the same airmass as the observations
of the program objects in order to correct for the presence of telluric lines in
the spectra. The $K$-band magnitude range for our sample stars is $\sim 7-12$,
resulting in exposure times in the range 600-2400 seconds in order to reach our
targeted S/N ratio of 50/1.  Spectra of ArNe arc lamps, taken before or after each program star, were used
for the wavelength calibration of our sample. Bias correction,
flat-fielding, spectral extraction, wavelength calibration, telluric feature
correction, and continuum normalization were all performed using standard IRAF
packages\footnote{IRAF is distributed by the National Optical Astronomy
Observatories, which is operated by the Association of Universities for Research
in Astronomy, Inc. under cooperative agreement with the National Science
Foundation.}.   

\section{Adopted Atmospheric Parameters and Synthetic Spectra}

Atmospheric parameters (\teff{}, \logg{}, and \metal{}) were estimated from
available optical and near-IR photometry, as well as from previously obtained
medium-resolution optical spectroscopy. Estimates of \teff{} are obtained from
measured V$-$K colors (taken from Beers et
al. 2007a, and references therein, as well as from the
2MASS Point Source Catalog; Skrutskie et al. 2006). The use of
near-IR photometry provides for a more accurate determination of
\teff{}, as the K band is less influenced by the presence of carbon features
than bluer bands. We used the Alonso et al. (1996) calibrations of \teff{} with V$-$K colors, as described in Paper I.  Surface gravities, \logg{}, have been been estimated based on
the Padova evolutionary tracks for metallicities \metal{}$=-2.5$ and \metal{}$=
-1.7$ (Girardi et al. 2000; Marigo et al. 2001).  Uncertainties in \teff{} and \logg{} are 100 K and 0.5 dex, respectively. The microturbulence is taken to be 2 km s$^{-1}$ for all stars.  This is consistent with previously-determined microturbulence values for giant CEMP stars (Johnson et al. 2007; Aoki et al. 2007). 

We have constructed two sets of synthetic spectral templates, covering
the optical and near-IR bands. Each set consists of 2000 synthetic spectra with
carbon-enhanced atmospheres generated with the MARCS code (Gustafsson
et al. 2008).  We used a previous generation of models here, as updated CEMP models were not available.  We do not, however, anticipate large differences in the models of spectra.  
The use of carbon-enhanced models is of particular importance
for cool CEMP stars, for which the atmospheric structure is significantly altered
by carbon (Masseron et al. 2006).  No 3D$\rightarrow$1D corrections have been applied to our estimates.  Recent studies of these effects on two hyper metal-poor stars (Collet et al. 2006) have revealed \ofe{} corrections of $\sim-0.8$ based on OH molecules, thereby lowering the measured abundance of oxygen.  However, the magnitude of such corrections is expected to decrease with increasing metallicity (Collet
et al. 2007).  As the metallicities of our targets range from $-1.0$ to $-2.8$, it is likely that the 3D effects are less severe.  In addition, more recent studies have been carried out (A. Ivanauskas, private communication) concerning the effects of convection on \ctwo{}, CH, CN, CO, NH, and OH molecules.  When compared to the results from Collet
et al. (2006), the magnitude of the corrections appear smaller.  

The synthetic
grid covers a range \teff{} = 4000 to 6000K , \logg{} = 0.0 to 5.0, \metal{} =
$-5.0$ to 0.0, and \cfe{} = 0.0 to +4.0. We adopt fixed
nitrogen abundances set to 0.5 dex less than the carbon abundances, which is 
roughly appropriate for CEMP stars. The CH and CN linelists used for the synthesis of the optical spectra are those
compiled by Plez (see Plez \& Cohen 2005). The
CO linelists used for the near-IR synthesis are taken from Kurucz (1993). The synthetic grids are then
degraded to match the resolving power of the observed spectra ($R=2000$ for the
optical spectra, $R=3000$ for the near-IR spectra).

The optical spectra are used for the determination of \metal{} and \ch{}. The
\ion{Ca}{2} K line is matched with the model spectra to estimate \metal{}, and the
\ctwo{} and \cn{} features are fit for the estimation of \ch{}
(see Paper I).  Our adopted atmospheric parameters,
as well as the derived \ch{} and \cfe{}, are listed in columns (2)-(6) of Table
1.

\section{Determination of \oh{}}

In order to determine \oh{} we employed the near-IR synthetic spectra
constructed from model atmospheres with carbon enhancements (see above). Each
synthetic spectrum covers the wavelength range 2.25$\mu$m$-$2.40$\mu$m. For all
combinations of these parameters, models were available with \ofe{} values of
0.0, $+0.4$, and $+0.8$.  

The first step in the determination of \ofe{} is to use the grid of synthetic
spectra in combination with the atmospheric parameters to create three
models with \ofe{} values of 0.0, $+0.4$, and $+0.8$. These models
are then used in order to estimate the \ofe{} of the program stars. Each star in
the sample has a previously estimated \teff{}, \logg{}, \metal{}, and \cfe{}.
Given the fact that four parameters are known, the routine begins with 48
models: 16 models for each value of \ofe{}. The 48 models are selected as having
the two closest values of \teff{}, \logg{}, \metal{}, and \cfe{} for each of the
three values of \ofe{}. Once selected, a linear interpolation over each
parameter is performed in order to create three final models, one for each value
of \ofe{}, with the known values of the four parameters. The three final models
for a set of typical atmospheric parameters are shown in Figure 1.
With the other parameters fixed, it is easy to see how a typical spectrum
changes with increasing oxygen abundance. Next, a linear interpolation over \ofe{} is
performed on the three final models, creating new model spectra with varying oxygen. It should be noted that while the interpolation was performed over \ofe{}, the metallicity is fixed, and therefore the oxygen varies as \oh{}.  In some cases, it was necessary to extrapolate beyond the boundaries of the model grid in order to find a good fit to the data.     

Due to the presence of four large CO band in the near-IR spectra, there were difficulties in fitting the continuum across the entire region.  It was important to fit the continua with a low-order function so that the depth of the absorption features were not artificially enhanced or lessened due to the continuum fit.  For this reason, the spectra of all stars in the sample were
trimmed around each of four \co{} bands, and a local continuum was fit for each band prior to spectral synthesis. With the use of the synthetic spectra, oxygen abundances were then estimated individually for
each of the four bands by minimizing $\chi^2$. A robust average using bisquare weighting of the four
separate values was taken as the final estimate of oxygen abundance, with an associated
robust estimate of the scatter in these values taken as the error of
determination.

Figure 2 shows the fitting technique applied to four stars from the
sample. Each row shows the four separate estimates of \oh{} for each star. Also plotted on each panel are synthetic spectra with \oh{} values that vary from the best-fitting spectra by $\pm 0.5$ dex.  Once
the robust average is applied the resulting estimates of \oh{} are $-1.2$ for
HE~0111$-$1346, $-0.5$ for HE~0519$-$2053, $-0.6$ for HE~1207$-$3156, and
$-1.2$ for HE~2339$-$0837.

\section{Results}

The distributions of \ofe{} versus three of the parameters used for their
determinations are shown in Figure 3.  The solid lines are linear fits to the data.  For the entire sample, there are no significant correlations of \ofe{} with \teff{}, \metal{}, or \cfe{}.  In the middle panel of Figure 3, it can be
seen that only one of the stars in our sample with \metal{} $<-2.5$ has a value
of \ofe{} less than $+1.0$. The bottom panel of Figure 3 shows the distribution of \ofe{}
versus \cfe{}, revealing that the majority of our sample (45 stars) have
\cfe{} $>+1.0$, and so meet the definition for CEMP stars given by
Beers
\& Christlieb (2005). The other stars exhibit carbon enhancements of \cfe{} $\ge
+0.5$, and thus are at least moderately enhanced in carbon. The average error in the
determination of \ofe{} for our entire sample is 0.4~dex.  We adopted a minimum
error for our \ofe{} estimates of 0.25~dex due to the influence of errors that arise
from the estimation of \teff{}, \logg{}, \metal{}, and \cfe{} (see Paper I for
details).  

In Figure 4, a carbon cut has been made, such that only the oxygen
abundances for those stars with \cfe{} $>+1.75$ are plotted against \metal{}.
The stars with the highest abundances of carbon exhibit some of the
lowest metallicities in our sample. This is not surprising, given that high
values of \cfe{} are often associated with lower metallicities. The fit to the
data shows a slight trend of increasing oxygen with decreasing metallicity.  The solid line is a least squares fit of \ofe{} as a function of \metal{}.  Only a marginally significant slope ($-0.616 \pm 0.314$) is found, hence the correlation is quite weak.  For comparison, the dashed line in this figure represents the the fit of
\ofe{} versus \metal{} for the carbon-normal stars from the Spite et al. (2005)
sample.  The \ofe{} estimates of the Spite et al. (2005) sample come from the forbidden
[\ion{O}{1}] $\lambda$6300 \AA{} line.  

\subsection{Statistical Comparison to High-Resolution Oxygen Estimates}

The present catalog of measured oxygen abundances available in the literature
for CEMP stars is still relatively small, due to the difficulty of obtaining
estimates from optical spectra, even at high spectral resolution. 
However, we can at least compare the regions of the \ofe{} vs. \metal{} parameter
space that are occupied by CEMP stars of various sub-classes, based
on previous high-resolution oxygen estimates, with those
from our present medium-resolution effort. 

Figure 5 shows \ofe{} for our entire sample, with different boxes indicating the
regions of parameter space occupied by several classes of CEMP stars. Sources
for the high-resolution data for different classes of CEMP stars can be found in
Masseron et al. (2010), and references therein. The majority of the stars in our
sample occupy regions of the diagram as CEMP stars that have confirmed,
high-resolution measurements of s-process-element enhancements. There is clearly
overlap with the region occupied by CEMP-r/s stars as well. Few of our stars
overlap with the region occupied by CEMP-no stars in the literature;
CEMP-no stars tend to be more metal-deficient than most of the stars in our sample. 

\subsection{High-Resolution Nitrogen Estimates}

For 13 of our program stars, high-resolution estimates of \nfe{} are available
from S. Lucatello (private communication) and/or Aoki et al. (2007). We selected
ten of these stars for which the available high-resolution estimates of \cfe{}
were within 0.5 dex of our medium-resolution estimates. The three that are
omitted from our analysis and discussion have associated high-resolution
\metal{} and/or \ch{} estimates that differ significantly from the medium-resolution estimates
of these species. We report values of high-resolution \cfe{}, \cfeh{}, using our estimates of \metal{} combined with the high-resolution \ch{}.
We report high-resolution estimates of \nfe{}, \nfeh{}, by combining
our estimates of \metal{} with the high-resolution values of \nh{}. Two of the
ten stars had high-resolution estimates both from Lucatello and
Aoki et al. (2007), and an average of the two available estimates was
taken.

The values of \teff{}, \logg{}, \metal{}, \ch{}, \cfe{}, \cfeh{},
\nfeh{}, \oh{}, \ofe{}, and \sigofe{} for our entire sample are
listed in Table 1.     

\section{Discussion}

We expect that the majority of CEMP stars in our sample have been polluted by a companion low-metallicity AGB star.  
In an AGB star, intershell oxygen is predicted to be closely related to intershell $^{12}$C, 
which in turn has a direct influence on the maximum $^{13}$C abundance.  
By studying these abundances, we can better understand the nature of the s-process, 
as the $^{13}$C($\alpha,n$)$^{16}$O reaction is a major neutron source for the s-process in AGB stars (Lugaro et al. 2003).  
According to current theory, oxygen production in AGB stars becomes increasingly significant with
decreasing metallicity (Herwig 2004, 2005; Campbell \& Lattanzio 2008;
Lau et al. 2009). Since the overabundance of oxygen is smaller at solar metallicities, lower metallicities are better suited for probing the primary production of oxygen.  In
addition, the observed abundance patterns of elements produced by the progenitor
is expected to depend on the mass (and metallicity) of the AGB star
(Herwig 2004;
Stancliffe \& Glebbeek 2008).  

In the following subsections we consider a number of issues that could potentially impact the interpretation of our measurements.  First, we compare the results of our CEMP stars to those of carbon-normal metal-poor stars (Section 6.1).  We then turn to recent literature on the abundance yields of low-metallicity AGB models to compare with our derived C, N, and O abundances (Section 6.2).  We also consider how dilution processes might be expected to alter our abundances (Section 6.3).  Finally, we address the sources of uncertainty in the AGB models, and how these may lead to altered abundances of C, N, and O (Section 6.4).     

\subsection{\ofe{} in Carbon-Normal and Carbon-Enhanced Metal-Poor Stars}

The linear fit to the full sample shown in the middle panel of Figure 3 is consistent
with the fit from the Spite et al. (2005) sample. However, most of the values of
\ofe{} from the carbon-normal sample are tightly distributed around a constant
value of $+0.7$.  Referring back to Figure 5, we see much more scatter in our values of \ofe{} for
the carbon-enhanced stars, with values reaching as high as $+2.0$. It can be
inferred that metal-poor stars, regardless of carbon-enhancement, commonly
exhibit enhancements of oxygen. However, when carbon enhancement is present,
additional oxygen enhancement can be expected as well, due to the fact that both
of these elements are enhanced by some of the same mechanisms. 

\subsection{C, N, and O:  Comparison with AGB Models}

For 10 of the 13 stars for which we report high-resolution \nfe{} estimates in
Table 1, Figure 6 shows \cfe{}, \nfe{}, and \ofe{} as a function 
of \metal{}.  One can notice an increase in the abundances of all three species
with decreasing metallicity.  The linear fits for each of the species
all have approximately the same slope, suggesting that abundances of carbon,
nitrogen, and oxygen are highly correlated with one another. With estimates of
\cfe{}, \nfe{}, and \ofe{}, we can compare our results to the predictions of
abundance yields due to AGB evolution as described by Herwig (2004). Both
carbon and oxygen are dredged up to the surface in AGB stars after thermal
pulses. The overabundance of these elements in low-metallicity AGB stars is
larger for lower initial masses (Herwig 2004), due to the fact that
the intershell mass is larger.  With a similarly large dredge up parameter and a smaller envelope, the enrichment of C and O in the envelope of a lower mass star is larger.  This can be seen in the top panel
of Figure 7, where the abundance predictions for C, N, and O
(Herwig 2004) are shown for five different AGB masses, ranging from 2
M$_\sun$ to 6 M$_\sun$, all with \metal{} $= -2.3$. The C, N, and O abundances
for ten stars with available \nfe{} in our sample are shown in a similar way in
the lower two panels of Figure 7. For most of the stars, the
relationship of carbon and oxygen is consistent with the models. The abundance
pattern for HE~0017$+$0055 is a very close match to an AGB star of about 3
M$_\sun$. The other nine stars exhibit some discrepancy with respect to nitrogen, which has been noted before for other CEMP stars with s-process element enhancement (Paper I; Sivarani et al. 2006). A previous effort to search for
metal-poor stars with large enhancements of nitrogen relative to carbon
yielded similar abundances (Johnson et al. 2007). All but five of the stars in our
sample are giants, and thus more mixing and dilution of any material transferred
from an AGB companion is expected, thereby resulting in such intermediate abundances of nitrogen (Paper I; Denissenkov \&
Pinsonneault 2008).  In addition, the possible occurence of H-ingestion flashes (HIF; Herwig 2003, 2005; Woodward
et al. 2008; Hajduk et al. 2005; Campbell \& Lattanzio
2008) could potentially enhance N in metal-poor stars.  

Nitrogen is an element that is very sensitive to CN cycling, and the C/N ratio indicates if the CN cycle has been activated partially, or whether mixing and thermodynamic conditions have been such that the CN cycle has reached equilibrium.  The latter is the case in hot-bottom burning (HBB), which is found in more massive AGB stars (Boothroyd et al. 1993).  In this case, the bottom of the convective envelope connects with the H-burning shell, allowing processing of envelope CN material in the H-shell.  Material lost at the surface is accordingly modified in its CN abundance ratios.  The limiting mass for the onset of HBB decreases with decreasing metallicity from $\sim 5$ M$_\sun$ at Z = 0.02 to $\leq 3$ M$_\sun$ for Z = 0.0 (Forestini
\& Charbonnel 1997; Siess et al. 2002).  Stars which experience HBB show a C/N ratio close to the equilibrium ratio ($<0.1$), as can be seen also in Figure 7 for the 4, 5, and 6 M$_\sun$ cases.  On the contrary, the 2 and 3 M$_\sun$ cases show very large C/N ratios.  Not even partial CN cycling has occurred in these models, and the small overabundance of N is entirely due to the first and second DUP (Herwig 2004).  Partial CN cycling at the bottom of low-mass giant envelope convection is a well observed and parametrically modeled feature (e.g. Denissenkov \&
VandenBerg 2003).  Whether the same process operates in AGB stars as well is currently debated (e.g. Karakas
et al. 2010).  

\subsection{Considering the Effects of Dilution}

Nitrogen estimates only exist for ten of our stars.  We compared the carbon and oxygen predictions from Herwig
(2004) to all of our CEMP stars.  None of the stars with available nitrogen estimates show the extremely low C/N ratio indicative of hot-bottom burning.  We therefore restrict the comparison of our sample to the 2 and 3 M$_\sun$ models from Herwig
(2004).  In Figure 8, we show all of our stars
that have \cfe{} $\geq +1.0$. The black square and the red triangle at the upper right of the figure are the model predictions for 2 M$_\sun$ and 3 M$_\sun$, respectively.  Clearly these predictions have higher carbon and oxygen estimates than our sample, but this is likely due to the fact that the effects of dilution are not considered.  We consider a parametric mixing model to test the effects of dilution of the accreted material from an AGB companion.  We chose a range of initial masses of the observed star ($0.5-0.9$ M$_\sun$) and accreted masses ($0.1-0.5$ M$_\sun$) and assumed complete mixing.  We set the maximum mass of the observed star at 1 M$_\sun$.  In order to consider only the AGB-phase contributions to carbon and oxygen, we subtracted off the likely contribution to carbon and oxygen arising from pre-star formation enhancements.  These contributions were chosen to be the average enhancements of carbon and oxygen from the Spite
et al. (2005) sample of unmixed metal-poor stars.  The subtracted enhancement of \cfe{} was $0.18\pm0.16$ dex and the subtracted enhancement of \ofe{} was $0.7\pm0.17$ dex.  The resulting predicted abundances based on this simple dilution experiment are shown as the black and the red lines in Figure 8.  Once the effects of dilution are considered, the AGB predictions of Herwig (2004) fall within the parameter space of our sample.  The magenta symbols are for the range of metallicity that is most consistent with the AGB models. For lower metallicities, the AGB
model values for \cfe{} and \ofe{} would be larger, and for higher metallicity,
they would decrease (Herwig
2004).    

\subsection{Uncertainties of the AGB Models}

The models of Herwig (2004) should be considered as rather conservative, standard predictions that suffer from the usual uncertainties associated mostly with convective mixing and mass loss.  We can obtain an indication of the order of magnitude of these uncertainties by comparing the Herwig (2004) models with those of other authors.  Comparing the 2 M$_\sun$ case with that of Cristallo
et al. (2009a), the carbon predictions agree very well, while the \nfe{} prediction of Cristallo
et al. (2009a) is about 0.2 dex higher and their O-overabundance prediction is 0.8 dex lower than that of Herwig (2004).  Karakas
(2010) provides a comparison between her Z = 0.0001, 2 M$_\sun$ yield predictions and that of Cristallo et al. (2009a) which shows that her C, N, and O yields are all approximately twice those of Cristallo et al. (2009a), which is easiest to be understood in terms of a lower mass loss rate in the Karakas (2010) yields.  Karakas (2010) provides a more in-depth discussion of model prediction differences from different authors.

In addition, we have to discuss uncertainties deriving from entirely alternative evolution scenarios.  Here, we should mention the possible occurrence of H-ingestion flashes (HIF), which have been introduced in Section 6.2.  These events are also referred to as proton-ingestion episodes or double He-shell flashes.  There are several uncertainties related to these combustion-type flashes in which protons are convectively mixed into the He-shell flash convection zone, and release energy on the dynamic time scale of convective flows.  

First, the occurrence of these events, in which the He-shell flash convection zone has to break through the entropy barrier associated with the H-shell, is more likely with lower CNO abundances in the envelope.  For example, an alpha-enhancement of the initial abundance composition without change of \metal{} will make the HIF less likely or even suppress it (Cristallo et al. 2009a).  If the stars are rotating (Meynet \& Maeder 2002), the CNO abundance may be enhanced during the core He-burning phase, which may also suppress the HIF (Herwig 2003).  

The second uncertainty in HIF predictions is the quantitative mixing and nucleosynthesis in a convective combustion regime that breaks some of the assumptions of mixing-length theory and 1D spherically symmetric stellar evolution (Herwig et al. 2010).  Keeping these significant uncertainties in mind, we nevertheless note that HIF models would predict the N signature of partial burning.  For example, considering Figure 6 in Cristallo et
al. (2009b), the prediction of a HIF model followed by maybe only a few pulses would correspond well to the CNO abundance patterns observed in most of our stars shown in Figure 7.  

\section{Conclusions}

We have used near-IR medium-resolution spectroscopy in order to estimate
\ofe{} for a sample of candidate carbon-enhanced stars selected
from the Hamburg/ESO Survey. This method of abundance analysis allows us to
obtain oxygen abundances accurate to about 0.4 dex. The use of four separate CO
features to estimate oxygen abundances from the near-IR spectra allows for more precise
estimates, based on a robust average of the independently determined fits.  A large spread of derived \ofe{} values are obtained for this sample,
ranging from near the solar value to as much as one hundred times greater.   

A comparison of our abundance determinations with high-resolution estimates was
carried out. The values of \ofe{} for our full set of 57 CEMP
stars largely fall within regions of parameter space occupied by the
high-resolution estimates of oxygen for other CEMP stars. We also found that the
majority of our stars have oxygen abundances that are consistent with known
CEMP-s and CEMP-r/s stars. Only a few stars could be considered CEMP-no stars,
based on the data compiled in Masseron
et al. (2010). This is likely due to
the fact that CEMP-no stars commonly have lower metallicities than most of the
stars in this sample.

Oxygen enhancements (on the order of \ofe{} $= +0.7$) have also been observed in
very metal-poor stars without significant carbon enhancement, indicating that
there were early oxygen-producing nucleosynthetic sites in the Galaxy
independent of any enhancement by AGB evolution. However, we find that the
\cfe{}, \ofe{}, and \nfe{} (when available) estimates follow the patterns from
Herwig (2004) closely enough that mass-transfer from a AGB companion is a
likely scenario for many of the stars in our sample, especially when the effects of dilution are considered.

Our measured carbon abundances always exceed the available high-resolution
\nfe{} abundances. If the origin of CNO abundance patterns comes from
hot-bottom-burning (HBB) in an intermediate mass (AGB) star, one would expect to
see elevated \nfe{} relative to \cfe{} and \ofe{}. This signature is not found
in our sample, but it has been suggested that other mechanisms, such as
cool-bottom-processing (Wasserburg et al. 1995; Denissenkov
\& VandenBerg 2003) or the occurrence of HIF,  can alter the levels of nitrogen enhancement.  

It is likely that the majority of CEMP stars in this sample will
turn out to be enhanced in neutron-capture elements.  Consistency of most of our program stars with the CEMP-s class, based both on
comparison to AGB models and existing high-resolution data, is expected
since that CEMP stars with s-process-element enhancement are the most commonly
observed type to date. 
However, recent chemical evolution models
(Cescutti \& Chiappini
2010) have revealed that the winds from massive, rapidly-rotating
metal-poor stars can result in a large scatter in the predicted abundances of C,
N, and O, presumably without the production of neutron-capture elements.
Therefore, we are currently unable to assign classification to this sample of CEMP stars.  High-resolution spectra of the stars in our sample will help clarify these
questions.

\acknowledgments C.R.K. and T.C.B. acknowledge partial support for this
work from grants AST 07-07776, as well as from PHY 02-15783 and PHY 08-22648; Physics
Frontier Center/{}Joint Institute or Nuclear Astrophysics (JINA),
awarded by the US National Science Foundation. S.R. and V.M.P. thank MSU/JINA, FAPESP, CNPq and Capes for financial support.  F.H. acknowledges funding through an NSERV Discovery Grant.

\begin{figure}
\epsscale{1.0}
\plotone{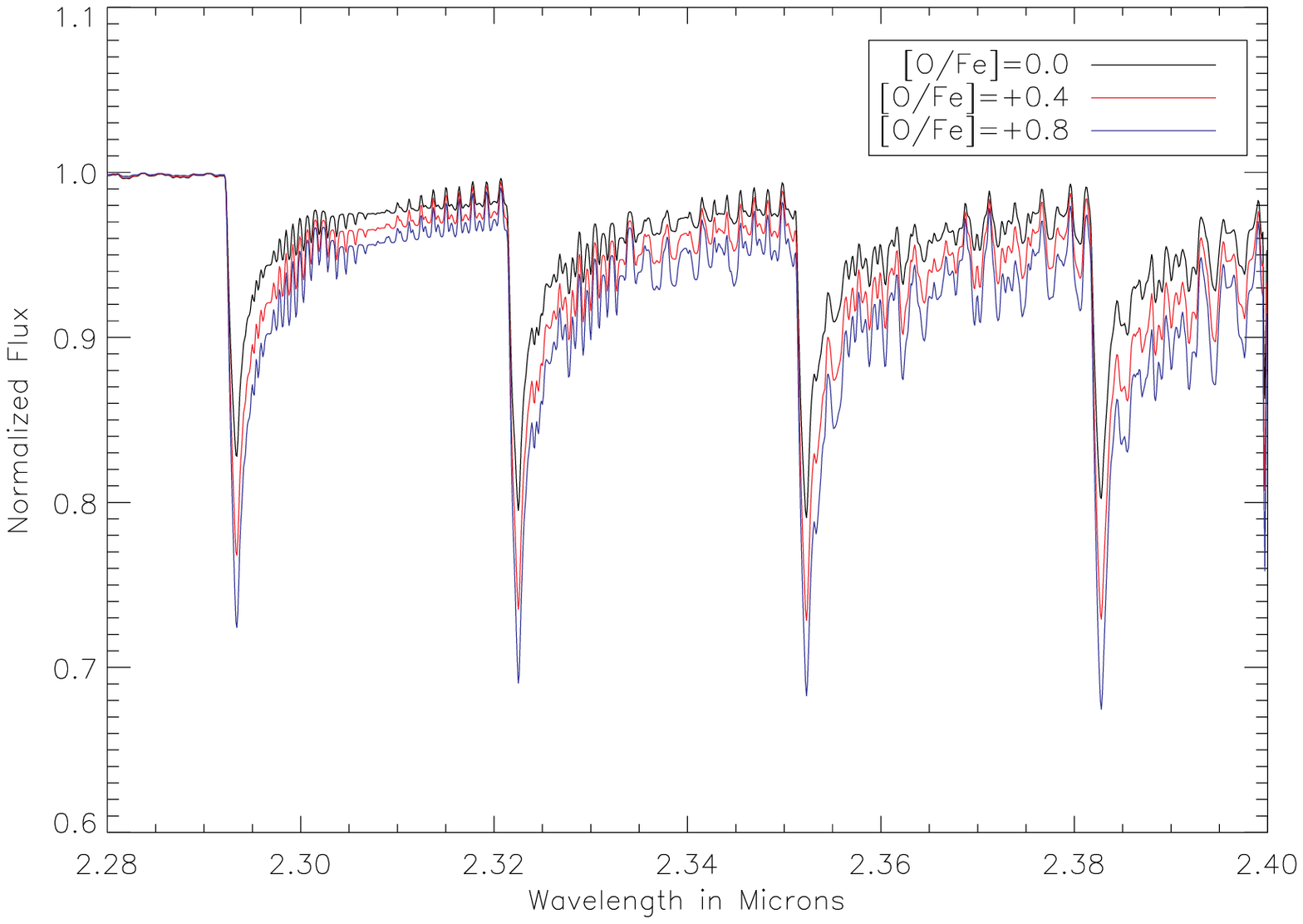}
\caption{Three synthetic spectra with different \ofe{} ratios.  Each spectrum has 
\teff{} of $4500$ K, \logg{} of 1.0, \metal{} of $-2.0$, and \cfe{} of $+1.0$.}
\label{models}
\end{figure}

\begin{figure}
\epsscale{1.0}
\plotone{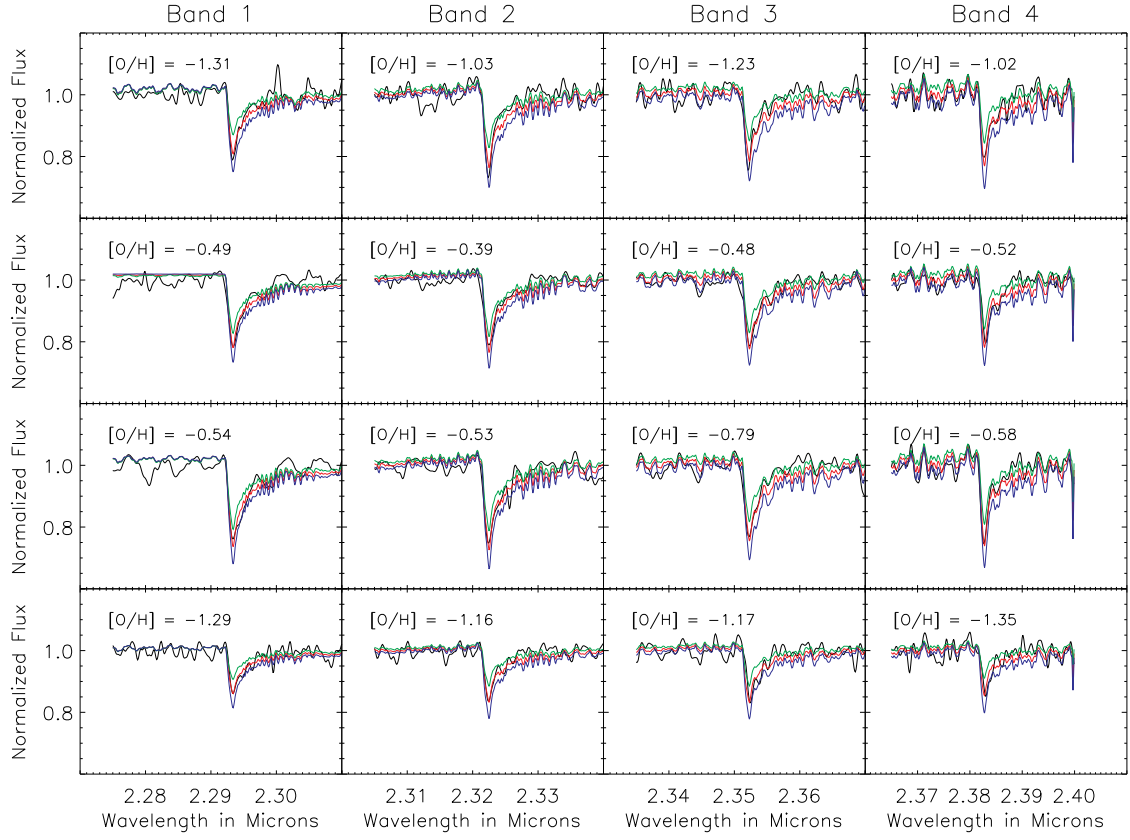}
\caption{Each row shows four estimates of \oh{} for a star from our sample: HE~0111$-$1346, HE~0519$-$2053, HE~1207$-$3156, and HE~2339$-$0837, respectively.  In each panel, the black lines are the data, the red lines are the best-fitting synthetic spectra, the green lines have \oh{} values of 0.5 dex lower than the best-fitting spectra, and the blue lines have \oh{} values of 0.5 dex higher than the best-fitting spectra.   
A robust average of the four separate estimates is taken as the 
final estimate of oxygen abundance for each star.  \textit{See text for details.}}
\label{bands}
\end{figure} 

\begin{figure}
\epsscale{0.65}
\plotone{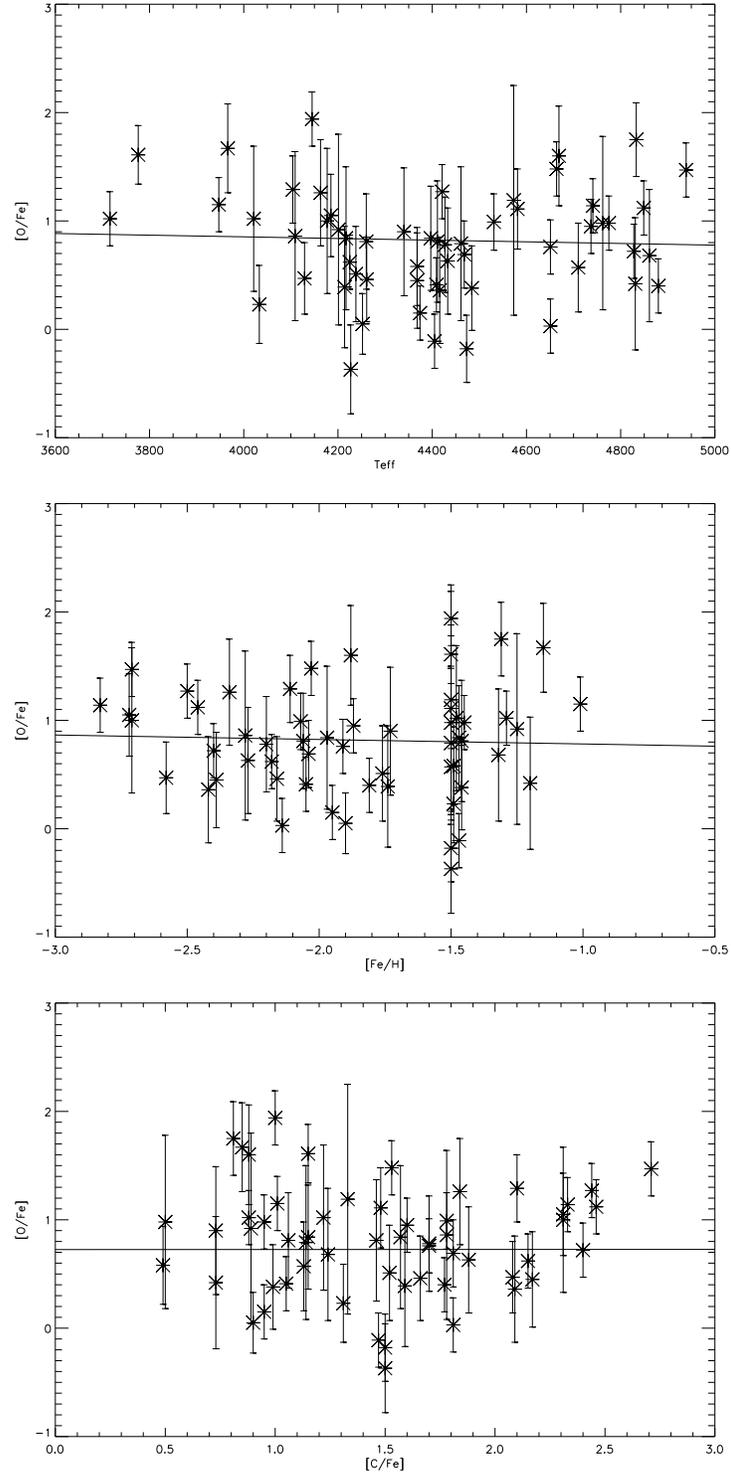}
\caption{\textit{Top panel:} \ofe{} vs. \teff{} for the entire sample.  \textit{Middle panel:} \ofe{} vs. \metal{} for the entire sample.  \textit{Bottom panel:} \ofe{} vs. \cfe{} for the entire sample.}
\label{ofevparams}
\end{figure} 

\begin{figure}
\epsscale{1.0}
\plotone{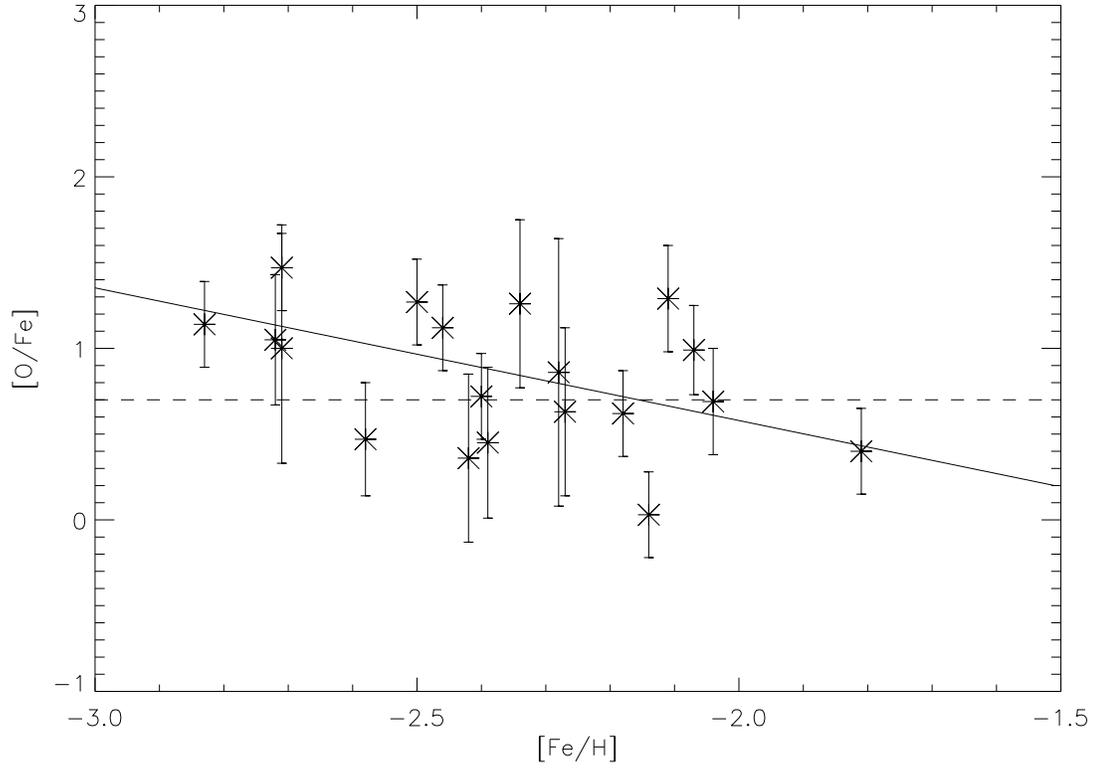}
\caption{\ofe{} vs. \metal{} for the stars with \cfe{} $\geq +1.75$. 
The dashed line represents the fit for the carbon-normal stars from the
Spite et al. (2005) sample of very metal-poor stars, while the solid
line is the best fit for our data.}
\label{hic}
\end{figure} 

\begin{figure}
\epsscale{1.0}
\plotone{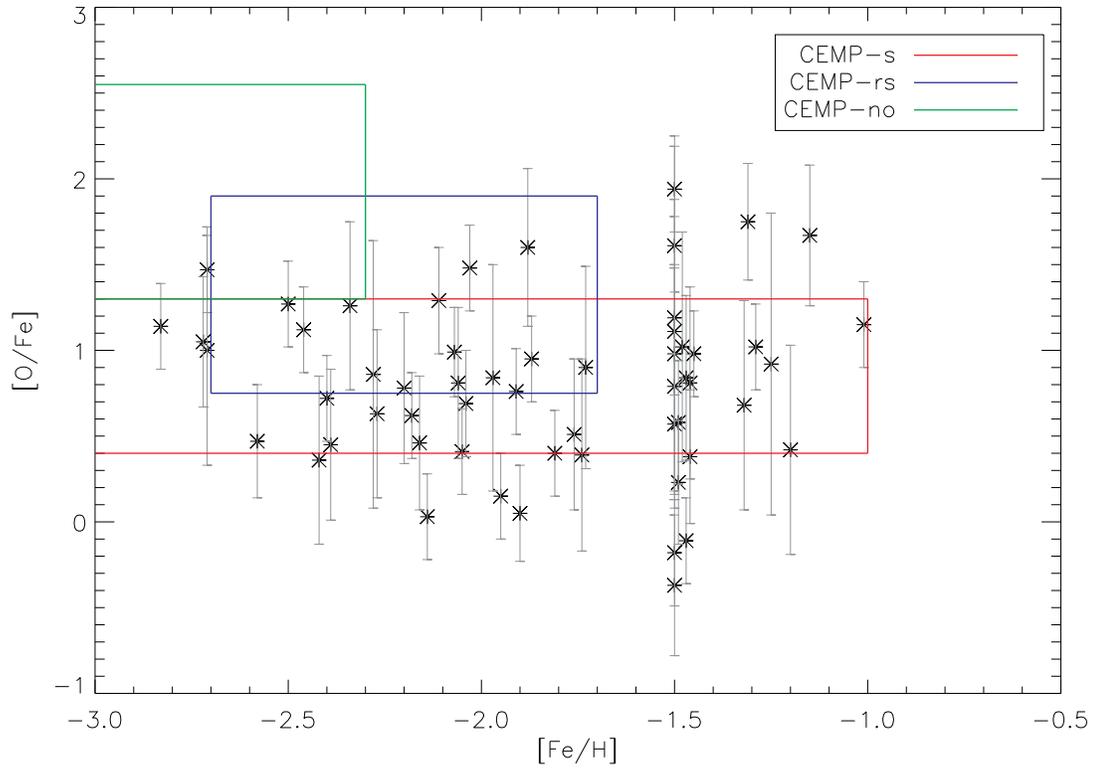}
\caption{\ofe{} vs. \metal{} for the entire sample.  The colored boxes show the regions occupied
by different types of \cemp{} stars found in Masseron et al. (2010).}
\label{ofebox}
\end{figure}

\begin{figure}
\epsscale{1.0}
\plotone{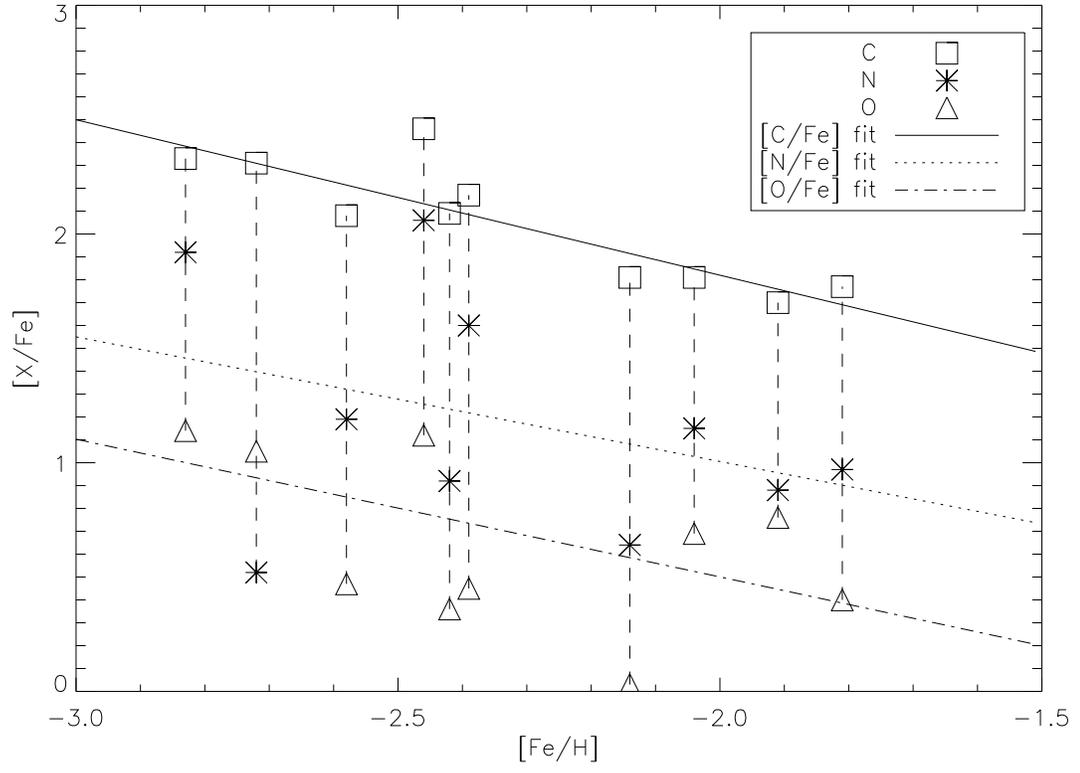}
\caption{C, N, and O abundances vs. metallicity for 10 stars from our sample.
Also shown are linear fits for these species.}
\label{cnometal}
\end{figure} 

\begin{figure}
\epsscale{0.60}
\plotone{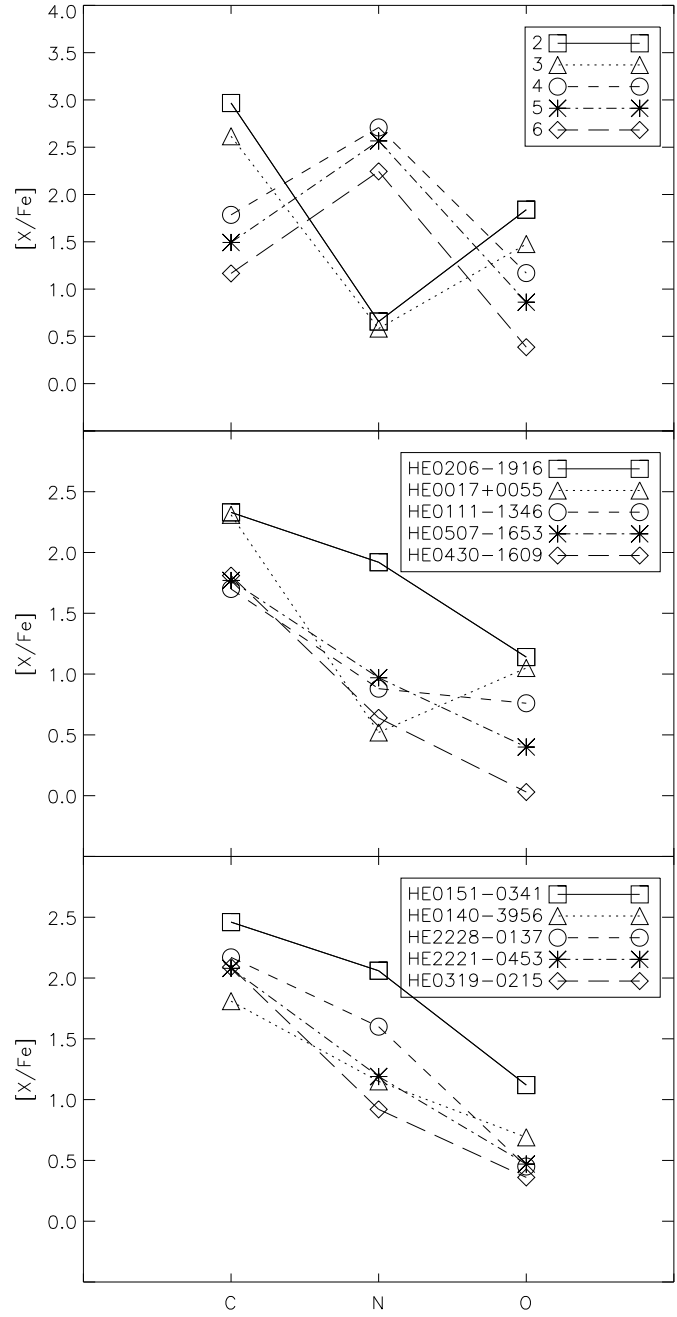}
\caption{\textit{Top panel:} Predicted abundances of C, N, and O for model AGB stars of different masses (in M$_\sun$) from Herwig (2004).
\textit{Middle and bottom panels:} Abundances of C, N, and O for 10 stars from
our sample.}
\label{herwigvosiris}
\end{figure}

\begin{figure}
\epsscale{1.0}
\plotone{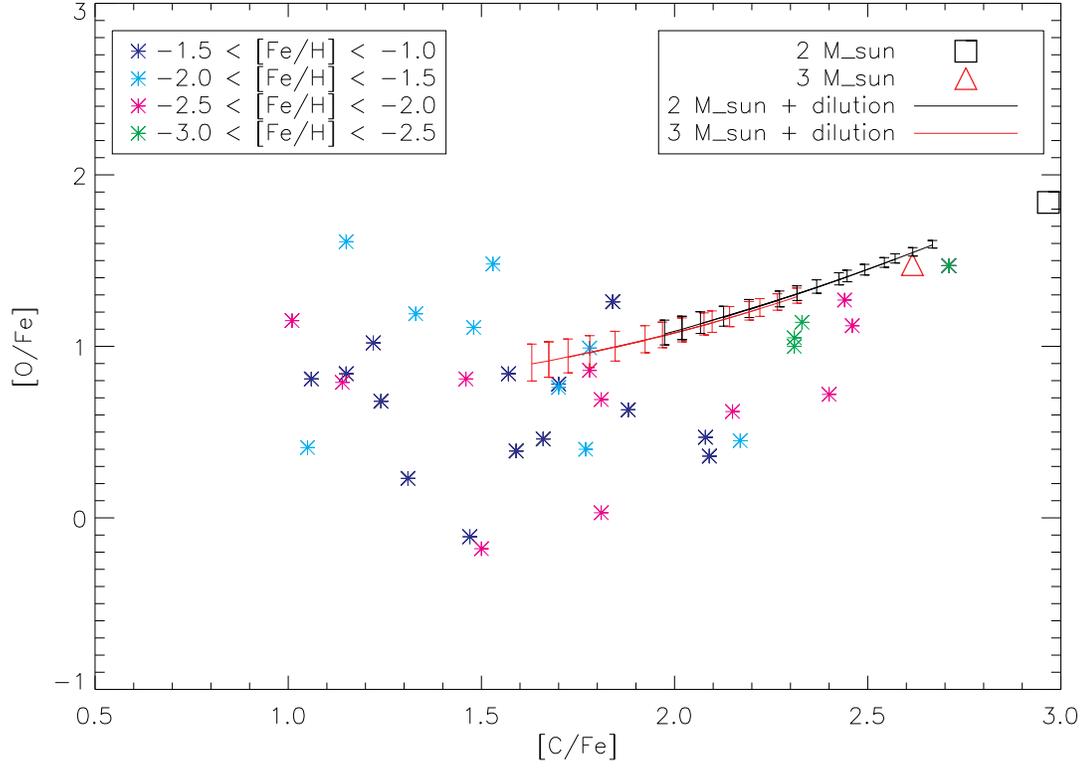}
\caption{\ofe{} vs. \cfe{} for all stars from our sample with \cfe $\geq +1.0$,
color-coded by their metallicity ranges, as noted in the legend.  The black square and red triangle mark the values of the abundances predicted by Herwig (2004) and the colored lines represent the effects of dilution once the AGB material is accreted onto the observed CEMP star.  \textit{See text for more details.}}
\label{ocherwig}
\end{figure}

\begin{deluxetable}{ccccrcccrrc}
\tablecaption{Estimates of Atmospheric Parameters and C, N, O Abundance Ratios}
\tablehead{Name & \teff{} (K) & \logg{} (cgs) & \metal{} & \ch{} & \cfe{} & \cfeh{} & \nfeh{} & \oh{} & \ofe{} & \sigofe{}}
\tablewidth{0pt}
\tabletypesize{\small}
\startdata
HE~0002$+$0053 & 4225 & 0.27 & $-$2.18 & $-$0.03 & 2.15 & \nodata & \nodata & $-$1.6 & 0.6 & 0.3 \\
HE~0010$-$3051 & 4177 & 0.17 & $-$2.71 & $-$0.40 & 2.31 & \nodata & \nodata & $-$1.7 & 1.0 & 0.7 \\
HE~0017$+$0055 & 4185 & 0.18 & $-$2.72 & $-$0.41 & 2.31 & 1.82 & 0.52 & $-$1.7 & 1.0 & 0.4 \\
HE~0033$-$5605 & 4021 & 0.00 & $-$1.48 & $-$0.26 & 1.22 & \nodata & \nodata & $-$0.5 & 1.0 & 0.7 \\
HE~0043$-$2433 & 4397 & 0.61 & $-$1.47 & $-$0.32 & 1.15 & \nodata &  \nodata & $-$0.6 & 0.8 & 0.5 \\
HE~0111$-$1346 & 4651 & 1.08 & $-$1.91 & $-$0.22 & 1.70 & 1.48 & 0.88 & $-$1.2 & 0.8 & 0.3 \\
HE~0120$-$5834 & 4828 & 1.62 & $-$2.40 & 0.00 & 2.40 & 1.79\tablenotemark{a} & 1.59\tablenotemark{a} & $-$1.7 & 0.7 & 0.3 \\
HE~0140$-$3956 & 4468 & 0.84 & $-$2.04 & $-$0.23 & 1.81 & 1.55 & 1.15 & $-$1.4 & 0.7 & 0.3 \\
HE~0151$-$0341 & 4849 & 1.42 & $-$2.46 & 0.00 & 2.46 & 2.16 & 2.06 & $-$1.3 & 1.1 & 0.3 \\
HE~0155$-$2221 & 4109 & 0.00 & $-$2.28 & $-$0.50 & 1.78 & \nodata & \nodata & $-$1.4 & 0.9 & 0.8 \\
HE~0206$-$1916 & 4741 & 1.23 & $-$2.83 & $-$0.50 & 2.33 & 2.42 & 1.92 & $-$1.7 & 1.1 & 0.3 \\
HE~0219$-$1739 & 4227 & 0.27 & $-$1.50 & 0.00 & 1.50 & 0.31\tablenotemark{a} & 0.31\tablenotemark{a} & $-$1.9 & $-$0.4 & 0.4 \\
HE~0251$-$2118 & 4710 & 1.16 & $-$1.50 & $-$0.37 & 1.13 & \nodata & \nodata & $-$0.9 & 0.6 & 0.4 \\
HE~0310$+$0059 & 4861 & 1.69 & $-$1.32 & $-$0.07 & 1.24 & \nodata & \nodata & $-$0.6 & 0.7 & 0.6 \\
HE~0314$-$0143 & 4201 & 0.22 & $-$1.25 & $-$0.36 & 0.89 & \nodata & \nodata & $-$0.3 & 0.9 & 0.9 \\
HE~0319$-$0215 & 4416 & 0.64 & $-$2.42 & $-$0.33 & 2.09 & 2.12 & 0.92 & $-$2.1 & 0.4 & 0.5 \\
HE~0330$-$2815 & 4411 & 0.64 & $-$1.46 & 0.00 & 1.46 & \nodata &  \nodata & $-$0.7 & 0.8 & 0.6 \\
HE~0359$-$0141 & 4340 & 0.54 & $-$1.73 & $-$1.00 & 0.73 & \nodata & \nodata & $-$0.8 & 0.9 & 0.6 \\
HE~0408$-$1733 & 4260 & 0.33 & $-$2.06 & $-$1.00 & 1.06 & \nodata & \nodata & $-$1.3 & 0.8 & 0.4 \\
HE~0417$-$0513 & 4669 & 1.22 & $-$1.88 & $-$1.00 & 0.88 & \nodata & \nodata & $-$0.3 & 1.6 & 0.5 \\
HE~0419$+$0124 & 4368 & 0.61 & $-$1.49 & $-$1.00 & 0.49 & \nodata & \nodata & $-$0.9 & 0.6 & 0.4 \\
HE~0429$+$0232 & 4409 & 0.63 & $-$2.05 & $-$1.00 & 1.05 & \nodata & \nodata & $-$1.6 & 0.4 & 0.3 \\
HE~0430$-$1609 & 4651 & 1.08 & $-$2.14 & $-$0.33 & 1.81 & 1.84 & 0.64 & $-$2.1 & 0.0 & 0.3 \\
HE~0439$-$1139 & 4833 & 1.62 & $-$1.31 & $-$0.50 & 0.81 & \nodata & \nodata & 0.4 & 1.8 & 0.3 \\
HE~0457$-$1805 & 4484 & 0.77 & $-$1.46 & $-$0.46 & 0.99 & \nodata & \nodata & $-$1.1 & 0.4 & 0.4 \\
HE~0458$-$1754 & 4374 & 0.56 & $-$1.95 & $-$1.00 & 0.95 & \nodata & \nodata & $-$1.8 & 0.2 & 0.3 \\
HE~0507$-$1653 & 4880 & 1.50 & $-$1.81 & $-$0.04 & 1.77 & 1.61 & 0.97 & $-$1.4 & 0.4 & 0.3 \\
HE~0518$-$1751 & 4252 & 0.32 & $-$1.90 & $-$1.00 & 0.90 & \nodata & \nodata & $-$1.9 & 0.1 & 0.3 \\
HE~0519$-$2053 & 4775 & 1.46 & $-$1.45 & $-$0.50 & 0.95 & \nodata & \nodata & $-$0.5 & 1.0 & 0.3 \\
HE~0547$-$4428 & 4217 & 0.25 & $-$1.97 & $-$0.40 & 1.57 & \nodata & \nodata & $-$1.1 & 0.9 & 0.7 \\
HE~1011$-$0942 & 3716 & 5.00 & $-$1.29 & $-$0.41 & 0.88 & \nodata & \nodata & $-$0.3 & 1.0 & 0.3 \\
HE~1023$-$1504 & 4421 & 0.66 & $-$2.50 & $-$0.06 & 2.44 & \nodata & \nodata &$-$1.2  & 1.3 & 0.3 \\
HE~1125$-$2942 & 3947 & 5.00 & $-$1.01 & 0.00 & 1.01 & \nodata & \nodata & 0.1 & 1.2 & 0.3 \\
HE~1145$-$0002 & 4033 & 0.00 & $-$1.49 & $-$0.18 & 1.31 & \nodata & \nodata & $-$1.3 & 0.2 & 0.4 \\
HE~1152$-$0355 & 4214 & 0.23 & $-$1.74 & $-$0.15 & 1.59 & \nodata & \nodata & $-$1.4 & 0.4 & 0.6 \\
HE~1152$-$0430 & 4573 & 0.93 & $-$1.50 & $-$0.17 & 1.33 & \nodata & \nodata & $-$0.3 & 1.2 & 1.1 \\
HE~1204$-$0600 & 4581 & 1.07 & $-$1.50 & $-$0.02 & 1.48 & \nodata & \nodata & $-$0.4 & 1.1 & 0.4 \\
HE~1207$-$3156 & 4664 & 1.23 & $-$2.03 & $-$0.50 & 1.53 & \nodata & \nodata & $-$0.6 & 1.5 & 0.3 \\
HE~1230$-$0230 & 3966 & 5.00 & $-$1.15 & $-$0.30 & 0.85 & \nodata & \nodata & 0.5 & 1.7 & 0.4 \\
HE~1238$-$0435 & 4433 & 0.68 & $-$2.27 & $-$0.39 & 1.88 & \nodata & \nodata & $-$1.6 & 0.6 & 0.5 \\
HE~1246$-$1510 & 3776 & 5.00 & $-$1.50 & $-$0.35 & 1.15 & \nodata & \nodata & 0.1 & 1.6 & 0.3 \\
HE~1255$-$2324 & 4405 & 0.69 & $-$1.47 & 0.00 & 1.47 & \nodata & \nodata &$-$1.6  & $-$0.1 & 0.3 \\
HE~1331$-$0247 & 4238 & 0.28 & $-$1.76 & $-$0.25 & 1.52 & \nodata & \nodata & $-$1.3 & 0.5 & 0.4 \\
HE~1410$-$0125 & 4427 & 0.66 & $-$2.20 & $-$0.50 & 1.70 & \nodata & \nodata & $-$1.4 & 0.8 & 0.4 \\
HE~1418$+$0150 & 4163 & 0.13 & $-$2.34 & $-$0.50 & 1.84 & \nodata & \nodata & $-$1.1 & 1.3 & 0.5 \\
HE~1428$-$1950 & 4531 & 0.85 & $-$2.07 & $-$0.28 & 1.78 & \nodata & \nodata & $-$1.1 & 1.0 & 0.3 \\
HE~1431$-$0755 & 4104 & 0.00 & $-$2.11 & $-$0.01 & 2.10 & \nodata & \nodata & $-$0.8 & 1.3 & 0.3 \\
HE~1524$-$0210 & 4145 & 5.00 & $-$1.50 & $-$0.50 & 1.00 & \nodata & \nodata & 0.4 & 1.9 & 0.3 \\
HE~2115$-$0522 & 4762 & 1.46 & $-$1.50 & $-$1.00 & 0.50 & \nodata & \nodata & $-$0.5 & 1.0 & 0.8 \\
HE~2145$-$1715 & 4461 & 0.77 & $-$1.50 & $-$0.36 & 1.14 & \nodata & \nodata & $-$0.7 & 0.8 & 0.7 \\
HE~2153$-$2323 & 4261 & 0.33 & $-$2.16 & $-$0.50 & 1.66 & 1.13\tablenotemark{a} & 0.93\tablenotemark{a} & $-$1.7 & 0.5 & 0.4 \\
HE~2200$-$1652 & 4473 & 0.84 & $-$1.50 & 0.00 & 1.50 & \nodata & \nodata & $-$1.7 & $-$0.2 & 0.3 \\
HE~2207$-$1746 & 4737 & 1.23 & $-$1.87 & $-$0.27 & 1.60 & \nodata & \nodata & $-$0.9 & 1.0 & 0.3 \\
HE~2221$-$0453 & 4129 & 0.04 & $-$2.58 & $-$0.50 & 2.08 & 2.08 & 1.19 & $-$2.1 & 0.5 & 0.3 \\
HE~2224$-$0330 & 4831 & 1.62 & $-$1.20 & $-$0.47 & 0.73 & \nodata & \nodata & $-$0.8 & 0.4 & 0.6 \\
HE~2228$-$0137 & 4368 & 0.61 & $-$2.39 & $-$0.22 & 2.17 & 1.90 & 1.60 & $-$1.9 & 0.5 & 0.4 \\
HE~2339$-$0837 & 4939 & 1.60 & $-$2.71 & 0.00 & 2.71 & \nodata & \nodata & $-$1.2 & 1.5 & 0.3 \\
\enddata
\tablenotetext{a}{These values are not used for analysis due to a large discrepancy in estimated \metal{}.} 
\end{deluxetable}

\end{document}